# Mechanical Twinning in Phosphorene

V. Sorkin[1*], Y.Q. Cai[1], D. J. Srolovitz[2], and Y.W. Zhang[1†]

[1]Institute of High Performance Computing, A*STAR, Singapore

[2]Departments of Materials Science and Engineering & Mechanical Engineering and Applied Mechanics, University of Pennsylvania, Philadelphia, PA 19104, USA

*We investigate the deformation and failure mechanisms of phosphorene sheet and nanoribbons under uniaxial tensile strain along the zigzag direction using the density functional tight-binding method. Surprisingly, twin-like deformation occurs homogenously across the phosphorene sheet; which significantly increases its failure strain. Vacancies within the sheet lead to the heterogeneous nucleation of twins at a lower critical strain which, subsequently, propagate across the entire sheet. Twin-like deformation always occurs heterogeneously in phosphorene nanoribbons (with or without vacancies). Propagation of the twins is interrupted by fracture which initiates along the ribbon edge. The underlying mechanism is bond breaking between the atoms within phosphorene puckers and simultaneous bond formation between the atoms in neighboring puckers. This unusual deformation behavior in phosphorene may be exploited in novel nano-electronic-mechanical applications.*

---

[*] Email: sorkinv@ihpc.a-star.edu.sg
[†] Email: zhangyw@ihpc.a-star.edu.sg

Phosphorene, a two-dimensional (2D) layer of black phosphorus, is a relatively new addition to the fast growing family of 2D materials [1–3]. Its semiconducting features [3–5] and unusual mechanical, thermal and transport properties [6–10] have attracted significant interest in view of its potential applications in flexible electronics [11, 12], thermal management [13, 14], batteries [15], p-n junctions [16], gas sensors [17] and solar-cells [18].

Although the deformation mechanisms and mechanical properties of graphene [19], boronitrene [20] and $MoS_2$ [21, 45] have been examined in details, similar studies on phosphorene have only recently begun [22–29]. For example, a recent study [30] showed that the deformation and failure of phosphorene are highly anisotropic. For loading along the armchair direction, three deformation stages were identified: an initial 'linear elastic' stage (closely related to the interactions between the adjacent nanoribbon puckers), followed by a 'bond rotation' stage (where the puckered structure of the phosphorene is flattened via bond rotation), and finally a 'bond stretching' stage (where P–P bonds are strained up to the rupture limit). Failure is caused by the breaking of the most highly strained bonds. For loading along the zigzag (ZZ) direction, however, the applied tensile strain leads to direct bond stretching and final bond breaking. Recent studies of the atomic structure, energetics and kinetics of point defects [22–26] found that (similar to graphene), phosphorene reconstructed around vacancies and the vacancy formation energy and the diffusion energy barrier in phosphorene were lower than those in all known 2D materials [22, 23, 26]. Vacancies were found to significantly degrade the mechanical properties, for example, by reducing the Young's modulus, fracture strain and strength (particularly along the ZZ-direction) [27–29].

Twinning is a fundamental plastic deformation mechanism commonly observed in many three-dimensional systems that competes with other deformation/failure mechanisms, such as dislocation slip and crack initiation and propagation [30]. Twinning should be even more common in 2D materials than in 3D materials since, at the nanoscale, initiation of plastic deformation generally requires higher stresses than their bulk counterparts. However, mechanical twinning (twin-like deformation) has never been reported in 2D materials. Hence, the fundamental question arises: Can mechanical twin-like deformation occur in 2D materials? The answer to this question is not only of significant scientific significance but also of potential impact on many potential novel applications of 2D materials.

Here, we report the first observations of twin-like deformation in phosphorene sheets and nanoribbons with or without vacancy defect strained along the zigzag direction (observed simulations employing density functional tight binding method). For perfect phosphorene sheet, twin-like deformation occurs homogeneously, causing a rotation of the zigzag orientation to the armchair direction, which significantly increases the failure strain. We further examine the effect of vacancy on the twin-like deformation and find that although the deformation becomes heterogeneous, twins can still propagate across the entire sheet. For phosphorene nanoribbons (with or without vacancies), however, twins also form heterogeneously, but they are unable to spread across the entire ribbon due to the interruption of crack formation and propagation. Our study also reveals the fundamental twin-like deformation and failure mechanisms in phosphorene. This unique twinning behavior of phosphorene may be used for novel nano-electronic-mechanical applications.

Here, we apply the tight-binding (TB) method [31] to examine the deformation and failure of phosphorene. The TB approach is chosen instead of DFT because of its computationally efficiency and accuracy, which make it readily applicable to relatively large nanoscale systems. More specifically, the density functional tight-binding (DFTB) method [32–34], which combines near-quantum mechanical accuracy and classical molecular dynamics efficiency, is the best approach for the intermediate scale systems (~500 atoms) subjected to large strains. Recently, DFTB has been applied in other studies of phosphorene sheets and nanostructures [35–43]. Our simulations employ the quantum mechanical simulation package "DFTB+"[32, 44].

In line with previous DFTB simulations [35, 41–43], a phosphorene sheet was constructed using a unit cell [18] of and vacancies [26] in phosphorene obtained from DFT calculations. Nanoribbons with zigzag edges were cut from the sheet and vacancies were positioned at the center. The dangling bonds at the nanoribbon edges were passivated by hydrogen. After geometry optimization, a uniform uniaxial tensile strain was applied quasi-statically along the ZZ-direction at 0 K. The tensile strain was increased incrementally, and at each step, the total energy was minimized (see [46] for details).

First, we investigated a defect-free phosphorene sheet subjected to tensile strain along the ZZ-direction. Figure 1(a) shows the strain energy per atom as a function of tensile strain. The initial configuration at zero strain ($\varepsilon=0$) is shown in the inset of Figure 1(a), where the top layer atoms are in red, and the bottom layer atoms are in blue. Note that the phosphorene puckers are oriented along the ZZ-direction (the x-axis). When the tensile strain is applied, the sample elongates along the ZZ-direction and contracts along the armchair AC (transverse) direction (the y-axis). The top (and bottom) layer atoms in the adjacent puckers are initially separated (see a pair of A and B atoms in the top layer in Figure 1(b)) by |AB|≈3.5 Å. Due to the transverse contraction, this distance decreases gradually and finally approaches the bond length limit (|AB|→δ=2.95 Å, see Figure 1(d)). On the other hand, the distance between the nearest atoms within the puckers (see the BC atom pair in the top layer in Figure 1(c)) increases (see Figure 1(d)). At a critical strain of ε=0.24 (see Figure 1(a,d)), the length of the BC atom bonds within the puckers exceeds a limit, causing these nearest neighbor bonds to break. This BC bond breaking is accompanied by the formation of new bonds between A and B atoms in adjacent puckers (see Figure 1(c)). This simultaneous bond breaking and forming process causes a homogeneous lattice rotation with respect to the initial lattice orientation, which may be described as homogeneous twin-like deformation (see Figure 1(c)). Through the twin-like deformation, the ZZ orientation of the phosphorene rotates into the AC orientation. The twin-like deformation leads to a drastic release of strain energy. Subsequently, the newly formed AC sheet is able to continue to strain, just like a pristine sheet strained along the x-direction. Hence, the twin-like deformation significantly increases the failure strain for phosphorene along the ZZ-direction. Since the phosphorene structures are identical before and after the structural transformation, the strain energy at the transition can be viewed as the energy barrier. The transition pathway is associated with adjacent puckers being pushed together (Poisson effect) to form new bonds, while the atoms within the puckers are pulled apart (bond breaking).

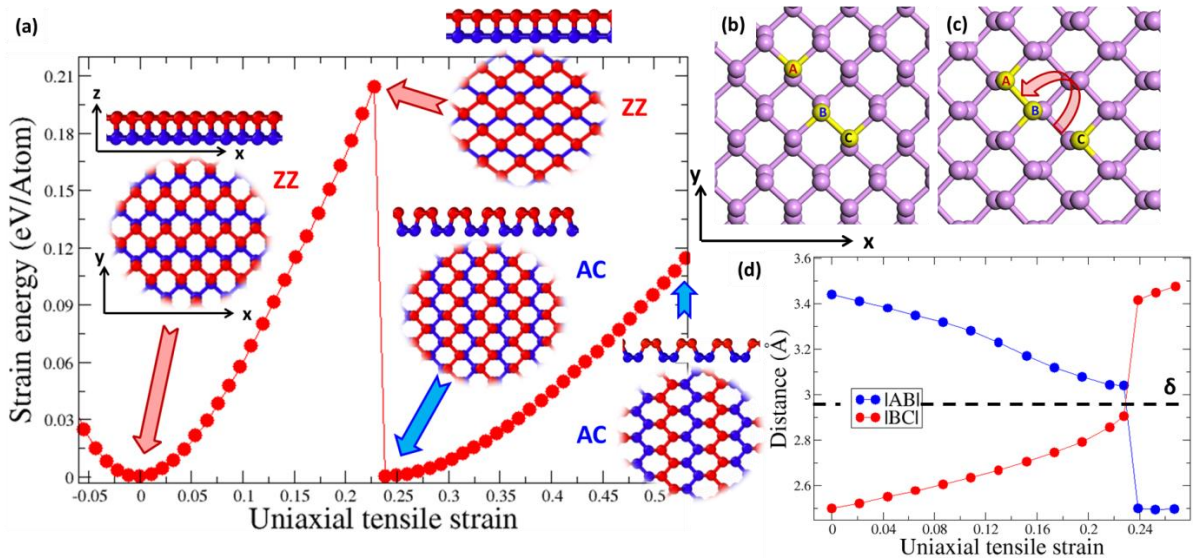

**Figure 1:** Structural transformation of defect-free phosphorene sheet under uniaxial tensile strain along the ZZ-direction. (a) Strain energy per atom as a function of tensile strain. The insets show the atomistic configurations/lattice orientation (side and top views) at the tensile strains indicated by the arrows. (b, c) Highlighted top layer atoms: A, B and C located in adjacent puckers. The images refer to the reference state ε=0 (b) and after the lattice reorientation at ε=0.24 (c). Arrow in (c) indicates the sense of the bond rotation. (d) The distances |AB| and |BC| between the nearest top-layer A, B and C atoms in adjacent puckers vs. tensile strain along the ZZ-direction (x-axis). The dashed line in (d) indicates the bond length limit.

Next, we introduced a vacancy (single or double) into the phosphorene lattice and examine how it affects the structural transformation observed upon straining in the ZZ-direction (see Figure 2). Interestingly, the twin-like deformation observed in pristine phosphorene also occurs in the defected sample, but at a much lower critical strain. Clearly, the presence of the vacancy is a stress concentrator, leading to heterogeneous twin-like deformation accompanied by a sequence of transitions (see Figure 2(a)). Here we describe the case of a (5|8|5) divacancy in detail, although similar results pertain to both a single vacancy and other divacancy variants [46].

The atomistic images in Figure 2(b-e)) illustrate the structural changes that occur upon application of the tensile strain. The atom color indicates the energy distribution within the sheet. At $\varepsilon=0$, the energies of the atoms located around the double vacancy are significantly higher than the average (see red atoms in Figure 2(b)). The bonds linking these atoms are considerably strained and, hence, rupture first at $\varepsilon=0.13$, triggering the transformation of the (5|8|5) divacancy into a (4|10|4) divacancy (see Figure 2(c)), and causing a small drop in the strain energy. A large energy drop follows at $\varepsilon=0.14$, at which the twinning propagates via a change in bond connectivity for many atoms around the vacancy (see Figure 2(d)). This is followed by the formation of a (5|7) edge dislocation at larger strain (indicated by circles in Figure 2(e)). The formed twin-like deformation domains gradually expand via bond rotations with increasing tensile strain. The twin propagation is completed by $\varepsilon=0.22$ although the initial bond connectivity of a small fraction of the atoms around the divacancy is preserved. At the final stage of the structural transformation, the (4|10|4) divacancy transforms back into a (5|8|5) vacancy (see Figure 2(e)). This new (5|8|5) divacancy is rotated by $\theta=90°$ from the original. Therefore, the entire sample is effectively rotated by $\theta=90°$ at $\varepsilon=0.22$ (except for the nucleation of the two edge dislocations and the lattice distortions caused by them). Clearly, the edge dislocation nucleation is a stress-relief mechanism.

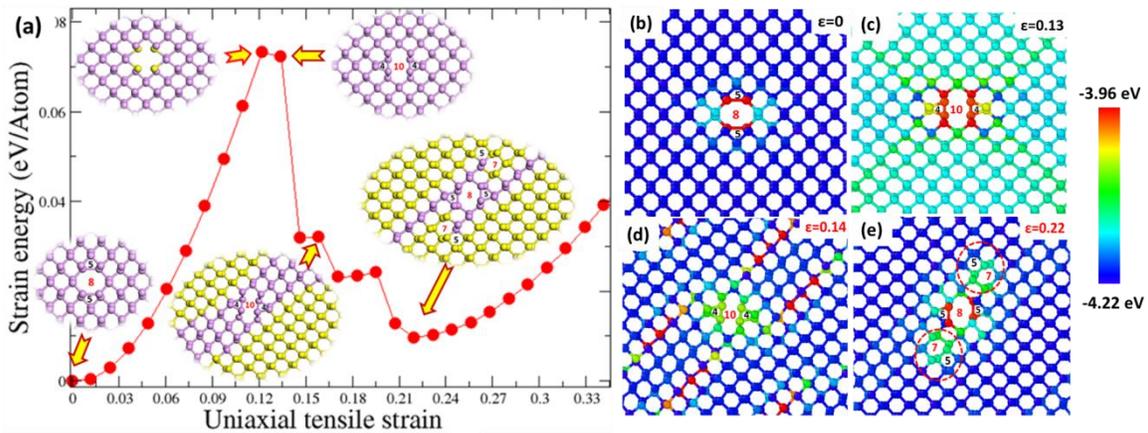

**Figure 2: Structural transformation in phosphorene sheet containing a divacancy upon uniaxial tensile strain along the ZZ-direction. (a) Strain energy per atom as a function of tensile strain. The inset shows atomistic configurations which are taken at the tensile strains indicated by arrows. Atoms in the twining domain are highlighted. (b-e) Spatial energy distribution: the atomistic snapshots are taken at the reference state (b), before (c) and after (d) the first transitions, and at the final stage (e) of the structural transformation. Color in (b)-(e) specifies energy per atom. Energy range is indicated by color bar on the right.**

Next, we examine the deformation and the failure mechanism of pristine and defected phosphorene nanoribbons with zigzag edges (see Figure 3). The atomic-scale images in Figure 3(b-e) show the tensile deformation of a perfect zigzag nanorribon (see also insets in Figure 3(a)). This uniform elastic deformation terminates abrubtly at ε=0.15 with the formation of a twin along the diagonal direction, leading to the first drop in the strain energy (see Figure 3(a-c)). This is accompanied by the formation of several pentagonal rings (encircled in Figure 3 (c)), where the twin emerges at the opposite edge from where it forms. Further straining induces the formation of another narrow twin along a different diagonal direction (see Figure 3 (e)). A further increase in the applied strain disrupts the continuing transformation by fracture, which initiates along one of the ribbon edges. Detailed examination of the failure process shows that the most strained bond at the pentagon-hexagon edge (see the red bond in Figure 3 (d)) ruptures, causing a sequence of bond breaking events (see Figure 3 (e)).

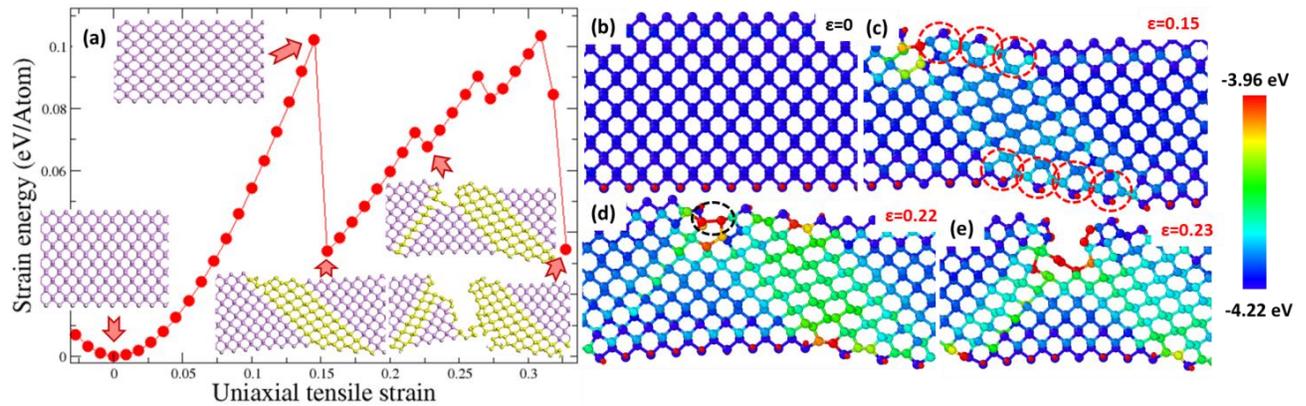

**Figure 3: Structural transformations in hydrogen-passivated phosphorene nanoribbons with zigzag edges under uniaxial tensile strain along the ZZ-direction: (a) Strain energy per atom of the defect-free zigzag nanoribbon as a function of tensile strain. Inserted atomistic configurations are taken at the tensile strains indicated by arrows. Atoms in the twining domain are highlighted. Phosphorous atoms are pink and hydrogen atoms are light-grey. (b-e) Set of atomistic snapshots taken in the reference state (b), at the first (c) and the second (d) drop in the strain energy, and at the final stage of the nanoribbon failure (e). Color in (b)-(e) specifies energy per atom. Energy range is indicated by color bar on the right.**

Finally, we examine the deformation and failure behavior of defected phosphorene nanoribbons (see Figure 4). Again, we focus on the (5|8|5) divacancy although similar results were found for a single vacancy and other divacancy variants [46]. The atomic-scale images in Figure 4 (b-e) show the main stages of tensile deformation (see also the insets in Figure 4(a)). The inhomogeneous structural transition starts as a sequence of bond rotations that initiates at the defect when the most stretched bonds around the divacancy rupture at ε=0.12 (see Figure 4 (a-c)). The bond connectivity changes simultaneously along the two diagonal directions (see Figure 4(c)). Bond rearrangement leads to the nucleation of a pair of pentagons at the edges (see Figure 4(c, d)). The divacancy undergoes a reconstruction: new bonds are formed between a pair of top layer atoms in adjacent puckers. The new (5|8|5) divacancy resembles the initial one, albeit rotated by θ=90° (see Figure 4(b, c)). Although the number of atoms within the twin gradually increases with applied strain, the structural transformation of the entire nanoribbon is preempted by fracture (as in the pristine nanoribbon). At a critical strain (ε=0.22), one of the most strained edge bonds (circled in Figure 4(d)) ruptures, causing a sequence of bond breaking events, which ultimately leads to the failure of the defected phosphorene nanoribbon.

Here, we present the first evidence of twin-like deformation in 2D materials. Twinning significantly increases the failure strain and serves as a stress relief mechanism. Since twin-like deformation leaves the final structure in the same pristine state as the original (except for a rotation), this suggests that this 2D material may be used for mechanical energy storage/switching/mechanical memory. We also investigated the possibility of reverse deformation of twinned samples subjected to uniaxial compression. Our DFTB calculations show that a completely reverse twinning transformation can be achieved in phosphorene nanosheets, if the buckling of the phosphorene sheet can be suppressed, for example, by using a short sample or using lateral supports [46]. Such reverse twin-like deformation may be used to create a mechanical switch and/or mechanical-based memory unit.

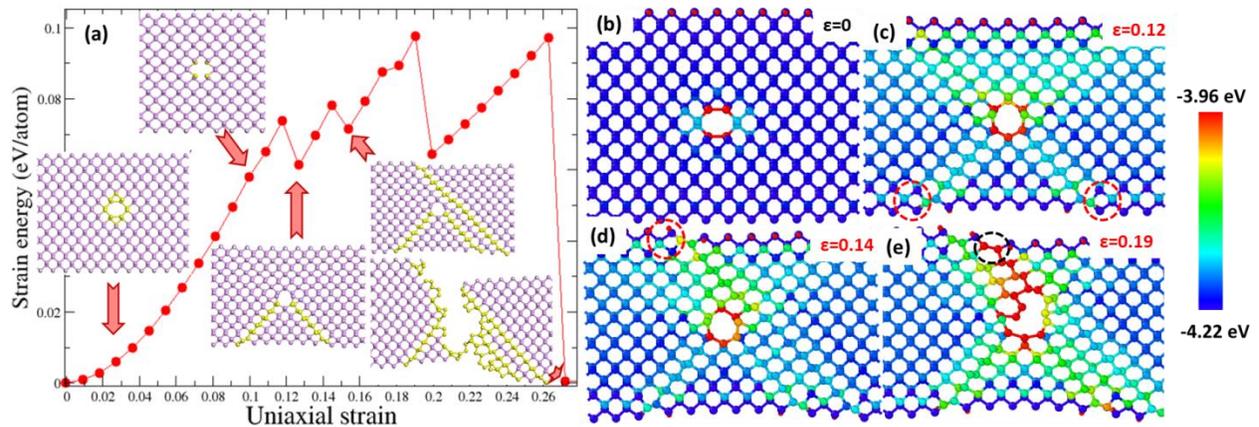

**Figure 4:** Structural transformations in hydrogen-passivated phosphorene nanoribbons with zigzag edges containing a divacancy under uniaxial tensile strain along the ZZ-direction. (a) The strain energy per atom as a function of tensile strain. The inset atomic configurations correspond to the tensile strains indicated by the arrows. Atoms in the twin are highlighted. (b-e) Set of atomic-scale images taken at the reference state (b), after the first (c) and the second (c) drop in the strain

energy, and at the final stage of the nanoribbon failure (e). Color in (b)-(e) specifies energy per atom. Energy range is indicated by color bar on the right.

**In conclusion, using DFTB calculations, we have investigated the deformation and failure of phosphorene sheets and nanoribbons subjected to uniaxial tensile strain along the ZZ-direction. We found that its peculiar anisotropic structure (with puckers oriented along the ZZ-direction) provides a pathway for the unusual twin-like deformation observed. Under the tensile strain, the bond length between the nearest neighbors within the puckers increases, while the distance between the next nearest neighbors in adjacent puckers decreases. At a critical strain, the bonds between the nearest neighbors break, while new ones form between next-nearest neighbors. As a result, the phosphorene lattice undergoes a homogeneous structural transformation/lattice rotation. The presence of vacancy leads to heterogeneous transformation initiated around the defect at a lower critical strain. With further tensile strain, the twinning domains expand, leading to the complete transformation of the lattice. Such a heterogeneous structural transformation always occurs in phosphorene nanoribbons. Bond rearrangement that occurs at the nanoribbon edges during twin-like deformation leads to the formation of pentagonal defects. The defects serve as the initiators of fracture that occurs through a bond rupturing sequence that ultimately leads to the failure of phosphorene nanoribbons. This unusual twinning behavior in phosphorene may promote the search for twin-like deformation in other 2D materials and seek their novel applications.**

This work was supported by the A*STAR Computational Resource Centre through the use of its high performance computing facilities, and in part by a grant from the Science and Engineering Research Council (152-70-00017). DJS gratefully acknowledges support as part of the Center for Computational Design of Functional Layered Materials (CCDM), an Energy Frontier Research Center funded by the U.S. Department of Energy, Office of Science, award DE-SC0012575.